\RequirePackage{latex/atlaslatexpath}
\documentclass[NOTE, atlasdraft=false, UKenglish, copyright=true]{atlasdoc}

\usepackage[subfigure]{atlaspackage}

\usepackage{atlasbiblatex}

\usepackage{atlasphysics}

\ifthenelse{\boolean{AtlasDraft}}{\usepackage[output=true, shift=true]{atlastodo}}{}

\graphicspath{{logos/}{figures/}}

\AtlasTitle{Recent measurements from the ATLAS experiment of Multi-Boson production processes at the LHC}

\AtlasVersion{1.0}

\AtlasAbstract{%

The high-energy proton-proton collisions at the Large Hadron Collider provide the ideal conditions to study the rare processes predicted by the Standard Model (SM) such as the production of multiple electroweak bosons. These processes involve the self-interactions of the gauge bosons through triple and quartic gauge couplings (TGCs and QGCs), in addition to interactions with the Higgs boson. Therefore, precision measurements of multi-boson final states probe the electroweak symmetry breaking mechanism and allow to search for deviations from the SM. Four recently published ATLAS results are summarised in this document.
The first two results are measurements of the production of di-boson final states in association with jets: the observation of $WW/WZ/ZZ$ production in semileptonic final states is described, and a measurement of the polarisation states of same-sign $W$ pairs is also discussed. The second set of measurements are related to the triple production of electroweak bosons: the first analysis reports the observation of $WW\gamma$ in the leptonic final state, while the second result presents the observation of triple vector boson production where at least one of the produced bosons is a $Z$ boson. All the measurements use datasets collected at a centre-of-mass energy of 13~TeV and corresponding to an integrated luminosity of 140~fb$^{-1}$.
}

\author{Diego Baron, on behalf of the ATLAS Collaboration \newline \newline Presented at the 32nd International Symposium on Lepton Photon Interactions at High Energies, Madison, Wisconsin, USA, August 25-29, 2025}

\AtlasRefCode{PROC-2025-104}

\AtlasCopyright{2025 CERN for the benefit of the ATLAS Collaboration.\newline
Reproduction of this article or parts of it is allowed as specified in the CC-BY-4.0 license.}

\hypersetup{pdftitle={ATLAS document},pdfauthor={Diego Baron, on behalf of the ATLAS Collaboration}}

\begin{document}

\maketitle



\section{Introduction}
\label{sec:intro}

In the Standard Model (SM), as a consequence of the non-abelian structure of the electroweak interactions, the electroweak bosons are subject to self-interactions. Multi-boson production processes receive contributions from self-interactions through triple and quartic gauge couplings (TGCs and QGCs). Therefore, the study of muti-boson production is key to the understanding of the SM electroweak sector including the couplings between the vector bosons and the Higgs boson. The high-energy conditions of proton-proton collisions at the Large Hadron Collider (LHC) \cite{Evans:2008zzb} are ideal to study these rare processes. The ATLAS detector \cite{PERF-2007-01} makes possible reconstructing and identifying the high-multiplicity of physics objects generated in multi-boson production. The ATLAS collaboration has performed several multi-boson measurements in the past \cite{STDM-2017-06,STDM-2022-06,STDM-2018-35,STDM-2017-19,STDM-2018-31,STDM-2018-59,STDM-2018-36,STDM-2018-32,STDM-2019-09,STDM-2018-33,STDM-2021-09,STDM-2019-17}. This work summarises the most recent ATLAS multi-boson results presented at the Lepton-Photon 2025 conference.

Four publications are summarised:
\begin{itemize}
	\item \textit{Electroweak diboson production in association with a high-mass dijet system in the semileptonic final states from $p$ $p$ collisions at $\sqrt{s} = 13$ TeV  with the ATLAS detector} \cite{STDM-2018-27}. This work reports the observation of $WW/WZ/ZZ$ production in association with a high-mass dijet system -- the analysis is optimised to enhance the contributions from the Vector Boson Scattering (VBS) process. The cross-section for VBS is measured in a fiducial region. The data are interpreted in the context of a dimension-8 effective field theory (EFT) to probe anomalous QGCs (aQGCs).
	\item \textit{Evidence for longitudinally polarized $W$ bosons in the electroweak production of same-sign $W$ boson pairs in association with two jets in pp collisions at $\sqrt{s} = 13$ TeV with the ATLAS detector} \cite{STDM-2019-26}. Evidence for the production of same-sign $W$ pairs where at least one $W$ boson is longitudinally polarised is reported. The most stringent constrain to date for the production of two longitudinally polarised same-sign $W$ bosons is set. 
	\item \textit{Observation of  $W^{+}W^{-}\gamma$ production in $p$ $p$ collisions at $\sqrt{s} = 13$ TeV and constraints on anomalous quartic gauge couplings} \cite{abcd}. The observation of $W^{+}W^{-}\gamma$ production is reported in final states containing one muon, one electron and a high-energy photon. The fiducial cross-section for this process is measured and constraints on 13 dimension-8 operators are set using the EFT framework.
	\item \textit{Observation of $VVZ$ production at $\sqrt{s} = 13$ TeV with the ATLAS detector} \cite{STDM-2020-08}.  The observation of the production of three massive vector bosons with at least one being a $Z$ boson is reported by selecting final states containing multiple leptons. The cross-section for the $WWZ$ process is measured and evidence for this individual production mode is found. Constraints on physics beyond the SM are also derived in the EFT framework by setting limits on Wilson coefficients for dimension-8 operators describing  aQGCs.
\end{itemize}

Section~\ref{sec:measurements} presents a brief summary of each measurement with every result having a dedicated sub-section. The conclusions are outlined in Section~\ref{sec:conclusion}.

\section{Multi-boson measurements}
\label{sec:measurements}

\subsection{WW/WZ/ZZjj production}
\label{sec:paper1}

The Vector Boson Scattering (VBS) process can take place in the SM through QGCs and also via the coupling between the Higgs and the electroweak bosons. The VBS process is of great interest because at high partonic centre-of-mass energy, the Higgs boson contribution keeps the VBS amplitude consistent with unitarity \cite{Lee:1977yc,Lee:1977eg}. At the LHC, the VBS process involves two partons radiating vector bosons that scatter. The VBS process generates a detector signature with two forward jets and two vector bosons (EWK $VVjj$) that can either decay hadronically or leptonically. In this work, events consistent with the EWK $VVjj$ process, in the channel where one of the bosons decays leptonically and the other hadronically (semileptonic channel) are studied. The semileptonic channel allows to access a higher boson transverse momentum ($\pt$) phase space than the pure leptonic final state. This is due to the enhancement in statistics provided by the larger branching fraction that vector bosons decaying into quarks have with respect to their leptonic decays.

Theories beyond the SM (BSM) predict anomalous quartic gauge couplings (aQGCs) \cite{Eboli:2003nq,Eboli:2006wa} or include new resonances \cite{Chang:2013aya, Espriu:2012ih}. These scenarios predict an enhancement of the VBS process at high vector boson $\pt$. This result uses the Eboli model \cite{Eboli:2006wa} to describe the possible BSM contributions to the EWK $VVjj$ processes in an EFT framework.

Three channels are explored in this analysis: 0-lepton corresponding to $Z\to\nu\nu$, 1-lepton for $W\to l\nu$ and 2-lepton  for $Z\to ll$ events. In all cases the other vector boson decays into a pair of quarks ($V\to qq$). In addition, the events are split according to how the $V\to qq$ decay is reconstructed. For events with low vector boson $\pt$, the quarks form two jets which can be resolved separately. However, in the high transverse momentum regime the two jet cones can merge and the $V\to qq$ decay is more efficiently reconstructed as a large radius jet (large-R jet). The inclusion of this merged category enhances the analysis sensitivity and it is the most sensitive category for the EFT interpretation. For large-R jets, substructure variables such as the jet mass are fed into a $W/Z$ boson tagging algorithm and events are categorised into high-purity and low-purity with respect to the scores given by the boson tagger.

\begin{figure}[h]
	\centering
	\includegraphics[width=0.6\linewidth]{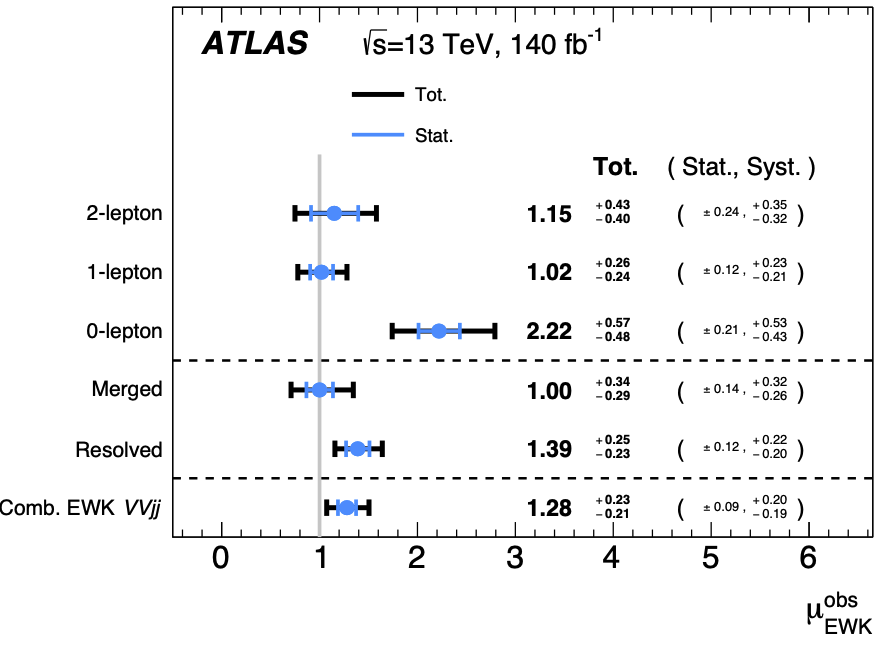}
	\caption{Signal stregth extracted from the combined fit, and from the independent fits using the channels with different lepton multiplicities and different jet reconstruction regimes. Both systematic and statistical uncertainties are displayed. Taken from \cite{STDM-2018-27}.}
	\label{Fig2}
\end{figure}

To extract the EWK $VVjj$ signal, a machine learning (ML) discriminant is trained in each channel to separate the EWK $VVjj$ process from other SM processes, leveraging the track multiplicity and kinematics of jets  The signal strength is extracted independently for the 0-, 1- and 2-lepton channels. In addition, the combination of all the channels is also used to perform the signal extraction. The fit results can be seen in Figure~\ref{Fig2}.

The EWK $VVjj$ process is observed (expected) with a significance of 7.4$\sigma$ (6.1$\sigma$) and the cross section for this process is measured in a fiducial region close to the detector and analysis acceptance. The obtained value is 29.2 $\pm$ 4.9 fb which is in agreement with the SM prediction. No significant deviations from the SM are found when searching for possible aQGCs using dimension-8 EFT operators contributions to the EWK $VVjj$ process.

\subsection{Same-sign $WWjj$ polarisation}
\label{sec:paper2}

The longitudinal polarisation states of the massive $W$ and $Z$ bosons are a result of the electroweak symmetry breaking mechanism in the SM. Hence, the study of the production of longitudinally polarised $W$ bosons in the VBS mode is crucial to test the Higgs mechanism. In this analysis, same-sign $W$ boson pairs where each individual boson decays into leptons ($W^{\pm}W^{\pm} \to l^{\pm}\nu l^{\pm}\nu$) are selected. The selected two-lepton combinations are: $e^{\pm}e^{\pm}$, $\mu^{\pm}\mu^{\pm}$ and $e^{\pm}\mu^{\pm}$. To enhance the contribution from the VBS process, two forward jets with a high dijet mass are also required completing the detector signature studied in this paper (EWK~$W^{\pm}W^{\pm}$jj)

\begin{figure}[h]
\centering
\subfigure[]{\includegraphics[width=0.329\textwidth]{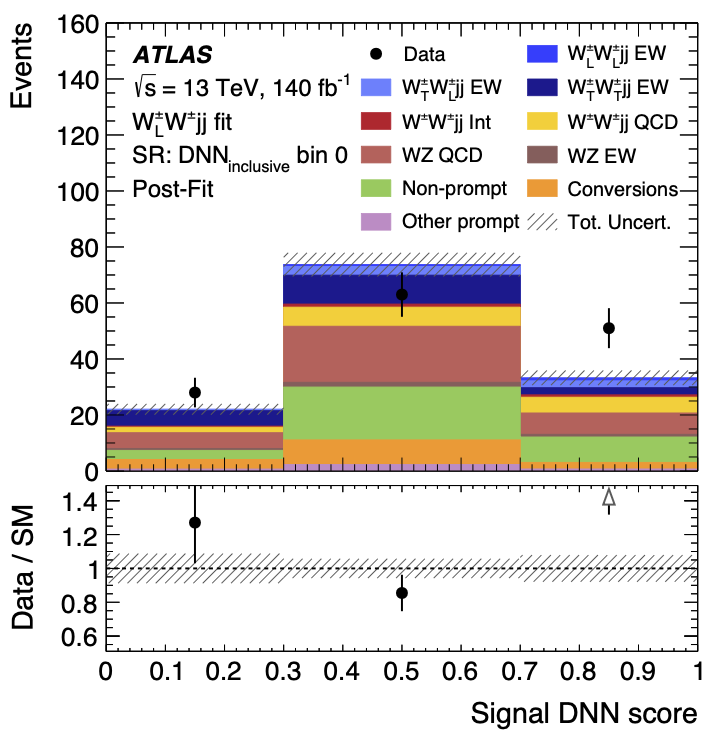}}
\subfigure[]{\includegraphics[width=0.329\textwidth]{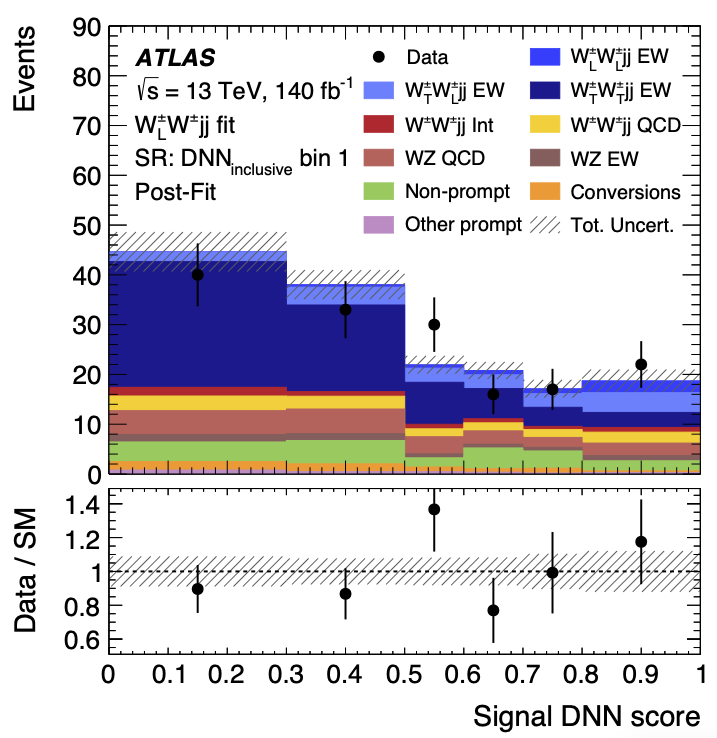}}
\subfigure[]{\includegraphics[width=0.329\textwidth]{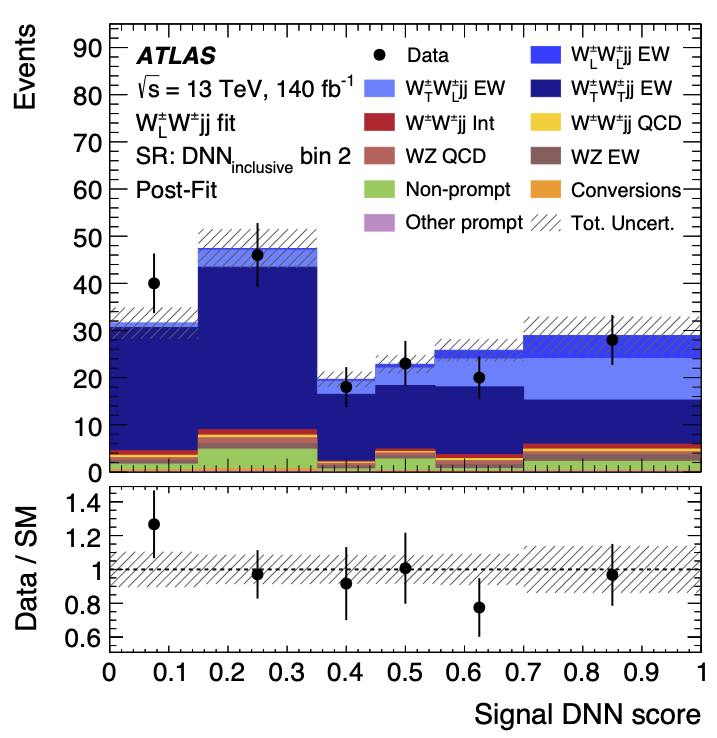}}
\caption{DNN scores of the algorithm trained to separate  $W^{\pm}_{L}W^{\pm}$ from  $W^{\pm}_{T}W^{\pm}_{T}$. Case (a), (b), (c) show events that have been classified by the inclusive DNN in bins with increasing background rejection power. Taken from~\cite{STDM-2019-26}.}
\label{Fig3}
\end{figure}

The polarisation states of the process are classified into three contributions: $W^{\pm}_{L}W^{\pm}_{L}$, $W^{\pm}_{L}W^{\pm}_{T}$ and $W^{\pm}_{T}W^{\pm}_{T}$, where  $W^{\pm}_{L}$ and $W^{\pm}_{T}$ denote longitudinally and transversely polarised W bosons, respectively. The polarisation vectors are defined in the centre-of-mass reference frame. The different polarisation states produce different kinematics for the jets and leptons, but there is not a single variable that can optimally separate the two contributions. Therefore, a set of kinematic variables are used to train deep neural networks (DNNs) to separate the different polarisation contributions. The $W^{\pm}_{L}W^{\pm}_{L}$ and $W^{\pm}_{L}W^{\pm}_{T}$ are extracted in separate fits since the number of available events is limited. The ML strategy is dived in two stages. First, a DNN is trained to separate the EWK~$W^{\pm}W^{\pm}$jj process from the backgrounds (DNN inclusive). In the second stage, two distinct DNNs are trained: one to separate $W^{\pm}_{L}W^{\pm}_{L}$ from $W^{\pm}W^{\pm}_{T}$ and $W^{\pm}_{L}W^{\pm}$ from  $W^{\pm}_{T}W^{\pm}_{T}$. The second stage DNN scores for the latter case can be seen in Figure~\ref{Fig3}.

The observed significance for the production of at least one longitudinally polarised $W$ boson is 3.3$\sigma$ with a measured cross section of 0.88 $\pm$ 0.30 fb which is in agreement with the SM. No significant excess consistent with the production of EWK $W^{\pm}_{L}W^{\pm}_{L}$jj is observed, hence a 95\% CL upper limit of 0.45 fb is measured for this process. 

\subsection{ $W^{+}W^{-}\gamma$  production}
\label{sec:paper3}

In the SM the vector bosons are predicted to interact with themselves via TGCs and QCGs. These  interactions contribute to processes where multiple vector bosons are produced. Therefore, measurements of triboson production provide stringent tests of the gauge structure of the SM and also probe potential new physics. In this analysis, events containing opposite charge $e\mu$ pairs, a high transverse momentum photon and missing energy ($WW\gamma\to e^\pm\mu^\mp\nu\bar{\nu}\gamma$) are selected to study the  $WW\gamma$ process.

\begin{figure}[h]
	\centering
	\includegraphics[width=0.5\linewidth]{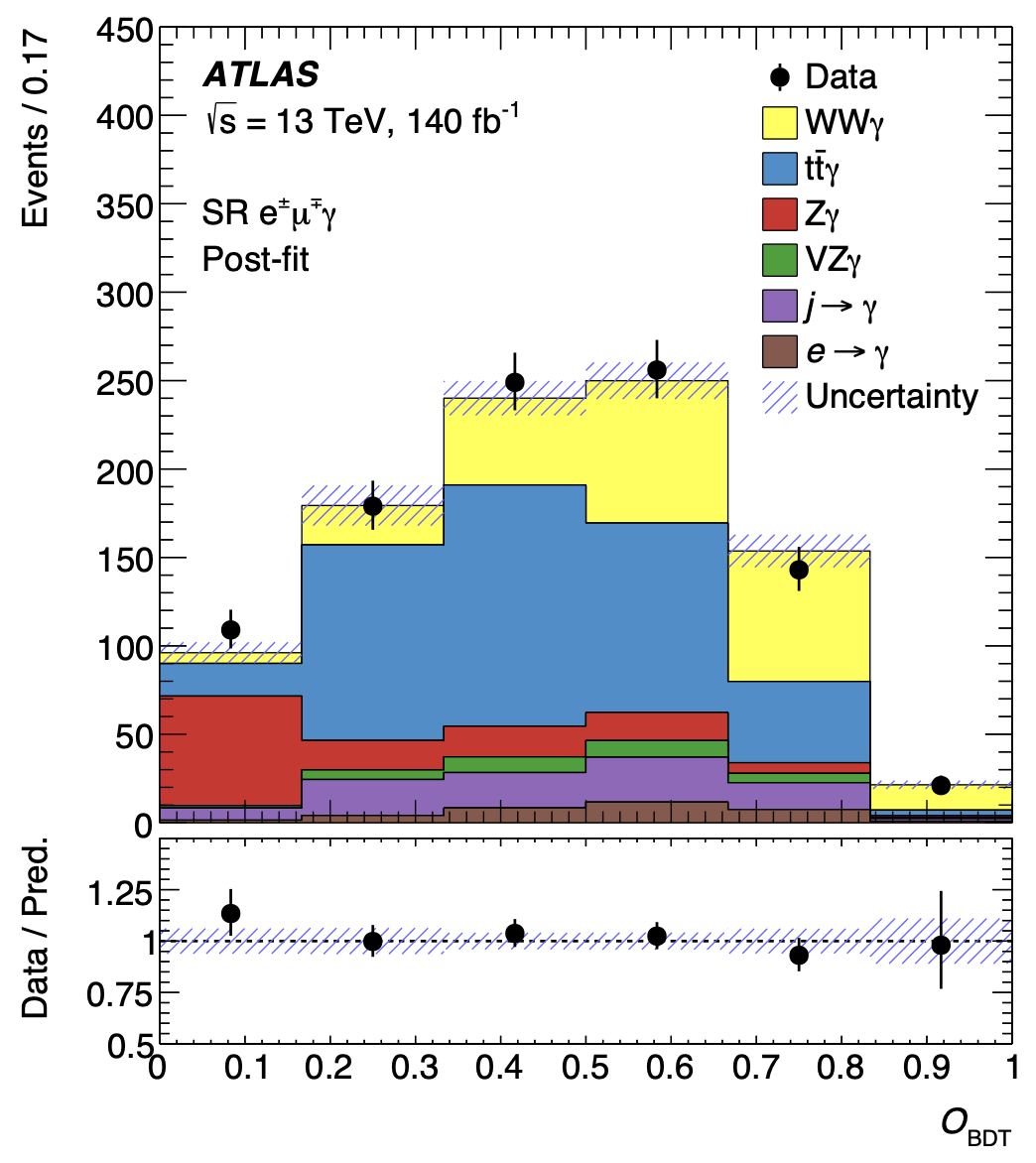}
	\caption{BDT score in the $WW\gamma$ signal region after applying all the normalisation factors extracted from the LH fit for the different processes. Taken from \cite{abcd}.}
	\label{Fig5}
\end{figure}

The main background contributions for this analysis come from events with prompt photons such as $t\bar{t}\gamma$ , $Z\gamma$ and $VZ\gamma$; and from events with non-prompt or misidentified photons ($e\to\gamma$ or $j\to\gamma$). Dedicated control regions are used to study and constrain these backgrounds. A boosted decision tree (BDT) is trained using kinematic variables that offer discrimination power between signal and background. MC Events from the signal region are used for the training. The BDT score distribution in the signal region can be seen in Figure~\ref{Fig5}.

The $WW\gamma$ signal strength is extracted by performing a simultaneous binned maximum-likelihood (LH) fit using the BDT distributions in the signal and control regions. The backgrounds normalisation are allowed to float in the fit. The observed significance of the $WW\gamma$ signal is 5.9$\sigma$. Additionally, the cross-section is measured in a fiducial region consistent with the detector and analysis acceptance; the value measured is 6.2 $\pm$ 1.0 fb which is consistent with the SM prediction. Anomalous QGCs are probed by using the EFT framework and 95\% CL limits are derived for 13 dimension-8 operators. These limits are the first ones derived from the $WW\gamma$ mode and are complementary to existent muti-boson results.

\subsection{ $VVZ$ production}
\label{sec:paper4}

As mentioned previously, studying triple vector boson production probes the SM TGCs and QGCs and potential new physics deviations. The production of three massive vector bosons also gets contributions from the Higgstrahlung process. ATLAS has performed a measurement of the $VVZ$ production targeting different channels: 3$l$ for $WWZ$ and $WZZ$, 4$l$ for $WWZ$, and at least 5$l$ for $WZZ$ and $ZZZ$.

\begin{figure}[h]
	\centering
	\subfigure[]{\includegraphics[width=0.329\textwidth]{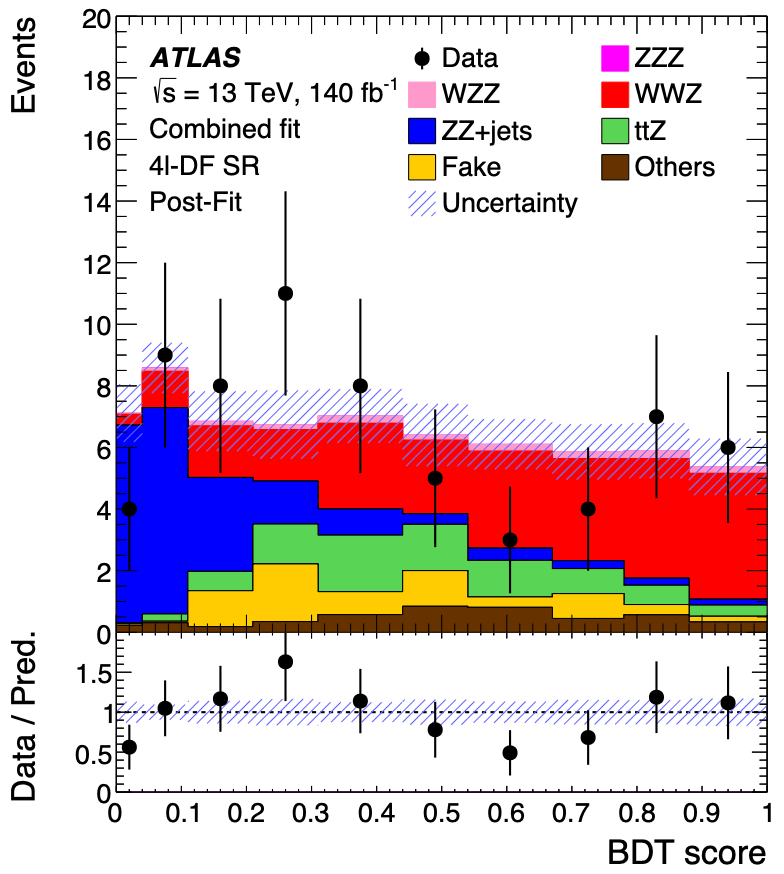}}
	\subfigure[]{\includegraphics[width=0.329\textwidth]{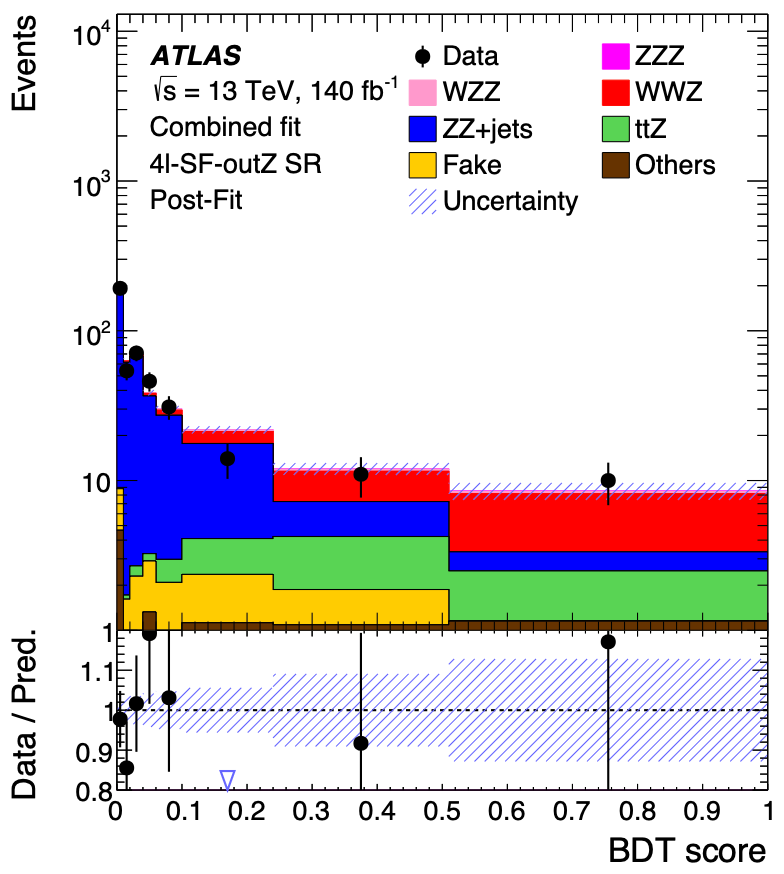}}
	\subfigure[]{\includegraphics[width=0.329\textwidth]{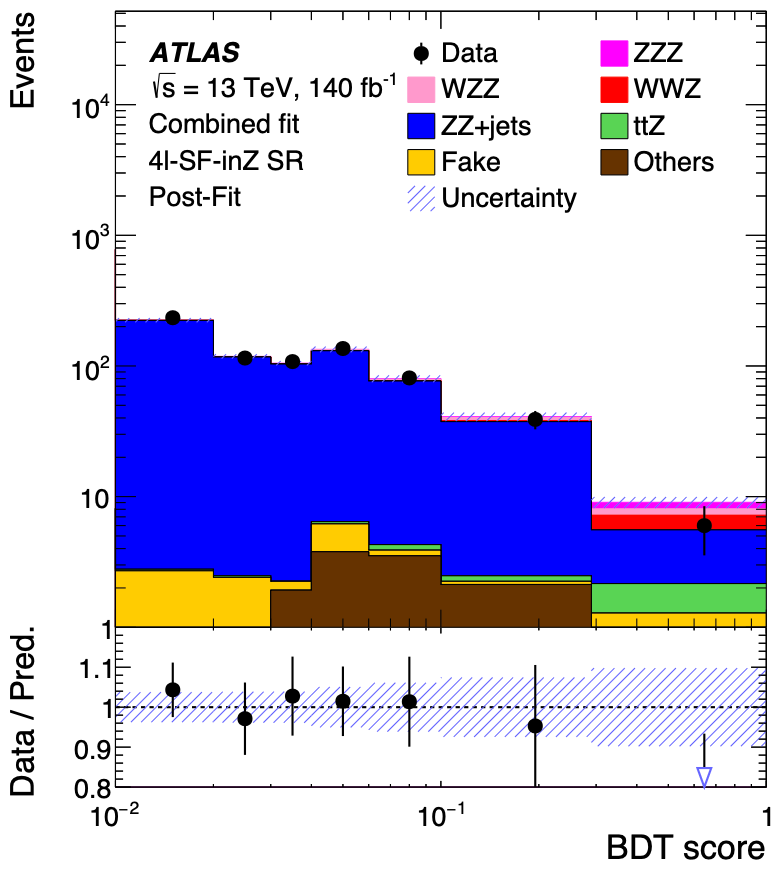}}
	\caption{Scores of the BDT algorithm trained to separate $VVZ$ from the background contributions in the 4$l$ channel. (a) shows a sample where the two leptons from the $W$ decays have different flavours (DF) and  same flavour (SF) for (b) and (c). The \textit{inZ} (\textit{outZ}) label refers to an event selection where the two $W$-leptons are required to have an invariant mass in (outside) a 20 GeV window from the $Z$ boson mass. Taken from~\cite{STDM-2020-08}.}
	\label{Fig7}
\end{figure}

As the multiplicity light-leptons increases, backgrounds where jets are misidentified as isolated leptons become more relevant compared to prompt leptons coming from V boson decays. Data-driven methods are used to estimate these backgrounds. The combined signal strength for the $VVZ$ process is extracted by performing a simultaneous profile likelihood fit in the seven signal regions and five control regions defined in this analysis. The distributions used in the fit are: the BDTs scores of individually trained (per lepton multiplicity) BDTs in the signal regions to discriminate between signal and background events, and the jet multiplicity distributions in the control regions. The BDTs scores for the 4$l$ signal regions can be seen in Figure~\ref{Fig7}. The observed significance of the $VVZ$ signal is 6.4$\sigma$ and the cross-section is found to be $660^{+93}_{-90}\text{(stat.)}^{+88}_{81}\text{(syst.)}$ fb, which is consistent with the SM prediction. The observed significance of $WWZ$ production is 4.4$\sigma$, representing evidcence for this process.

When the results are interpreted via the EFT framework to look for deviations from the SM, constraints are derived for the Wilson coefficients corresponding to dimension-8 operators describing aQGCs. These constraints are comparable to results already published by the ATLAS collaboration in the $W\gamma$jj final state.

\section{Conclusion}
\label{sec:conclusion}

This work summarises four recent multi-boson production measurements from the ATLAS collaboration using proton-proton collisions at a centre-of-mass energy of 13 TeV corresponding to an integrated luminosity of 140~$\text{fb}^{-1}$. These results expand the comprehensive muti-boson measurements program conducted by the ATLAS collaboration. Key observations include the electroweak production of di-bosons in association with a high-mass dijet system and the production of $WW\gamma$. Furthermore, evidence was found for the production of same-sign $W$ boson pairs with at least one longitudinally polarized boson, and for the $WWZ$ production process as part of the overall observation of $VVZ$ production. All cross-section measurements were found to be in agreement with Standard Model predictions, and no significant deviations were observed in the searches for anomalous quartic gauge couplings, leading to constraints on dimension-8 effective field theory operators.





\clearpage
\printbibliography

@Article{STDM-2018-27,
	author         = "{ATLAS Collaboration}",
	title          = "{Electroweak diboson production in association with a high-mass dijet system in semileptonic final states from \(pp\) collisions at \(\sqrt{s} = 13\,\text{TeV}\) with the ATLAS detector}",
	year           = "2025",
	reportNumber   = "CERN-EP-2025-050",
	eprint         = "2503.17461",
	archivePrefix  = "arXiv",
	primaryClass   = "hep-ex",
}

@Article{STDM-2019-26,
	author         = "{ATLAS Collaboration}",
	title          = "{Measurements and interpretations of \(W^{\pm}Z\) production cross-sections in \(pp\) collisions at \(\sqrt{s} = 13\,\text{TeV}\) with the ATLAS detector}",
	year           = "2025",
	reportNumber   = "CERN-EP-2025-145",
	eprint         = "2507.03500",
	archivePrefix  = "arXiv",
	primaryClass   = "hep-ex",
}

@Article{STDM-2020-08,
	author         = "{ATLAS Collaboration}",
	title          = "{Observation of \(VVZ\) production at \(\sqrt{s}=13\) TeV with the ATLAS detector}",
	year           = "2024",
	reportNumber   = "CERN-EP-2024-321",
	eprint         = "2412.15123",
	archivePrefix  = "arXiv",
	primaryClass   = "hep-ex",
}

@Article{PERF-2007-01,
    author         = "{ATLAS Collaboration}",
    title          = "{The ATLAS Experiment at the CERN Large Hadron Collider}",
    journal        = "JINST",
    volume         = "3",
    year           = "2008",
    pages          = "S08003",
    doi            = "10.1088/1748-0221/3/08/S08003",
    primaryClass   = "hep-ex",
}

@Article{STDM-2017-06,
    author         = "{ATLAS Collaboration}",
    title          = "{Observation of Electroweak Production of a Same-Sign \(W\) Boson Pair in Association with Two Jets in \(pp\) Collisions at \(\sqrt{s} = 13\,\text{TeV}\) with the ATLAS Detector}",
    journal        = "Phys. Rev. Lett.",
    volume         = "123",
    year           = "2019",
    pages          = "161801",
    doi            = "10.1103/PhysRevLett.123.161801",
    reportNumber   = "CERN-EP-2019-008",
    eprint         = "1906.03203",
    archivePrefix  = "arXiv",
    primaryClass   = "hep-ex",
}

@Article{STDM-2017-19,
    author         = "{ATLAS Collaboration}",
    title          = "{Observation of electroweak production of two jets and a \(Z\)-boson pair}",
    journal        = "Nature Phys.",
    volume         = "19",
    year           = "2023",
    pages          = "237--253",
    doi            = "10.1038/s41567-022-01757-y",
    reportNumber   = "CERN-EP-2020-016",
    eprint         = "2004.10612",
    archivePrefix  = "arXiv",
    primaryClass   = "hep-ex",
}

@Article{STDM-2018-31,
    author         = "{ATLAS Collaboration}",
    title          = "{Fiducial and differential cross-section measurements of electroweak \(W\gamma jj\) production in \(pp\) collisions at \(\sqrt{s} = 13\,\text{TeV}\) with the ATLAS detector}",
    journal        = "Eur. Phys. J. C",
    volume         = "84",
    year           = "2024",
    pages          = "1064",
    doi            = "10.1140/epjc/s10052-024-13311-6",
    reportNumber   = "CERN-EP-2024-048",
    eprint         = "2403.02809",
    archivePrefix  = "arXiv",
    primaryClass   = "hep-ex",
}

@Article{STDM-2018-32,
    author         = "{ATLAS Collaboration}",
    title          = "{Measurement and interpretation of same-sign \(W\) boson pair production in association with two jets in \(pp\) collisions at \(\sqrt{s} = 13\,\text{TeV}\) with the ATLAS detector}",
    journal        = "JHEP",
    volume         = "04",
    year           = "2024",
    pages          = "026",
    doi            = "10.1007/JHEP04(2024)026",
    reportNumber   = "CERN-EP-2023-221",
    eprint         = "2312.00420",
    archivePrefix  = "arXiv",
    primaryClass   = "hep-ex",
}

@Article{STDM-2018-33,
    author         = "{ATLAS Collaboration}",
    title          = "{Observation of \(W\gamma\gamma\) triboson production in proton--proton collisions at \(\sqrt{s} = 13\,\text{TeV}\) with the ATLAS detector}",
    journal        = "Phys. Lett. B",
    volume         = "848",
    year           = "2024",
    pages          = "138400",
    doi            = "10.1016/j.physletb.2023.138400",
    reportNumber   = "CERN-EP-2023-037",
    eprint         = "2308.03041",
    archivePrefix  = "arXiv",
    primaryClass   = "hep-ex",
}

@Article{STDM-2018-35,
    author         = "{ATLAS Collaboration}",
    title          = "{Measurements of electroweak \(W^{\pm}Z\) boson pair production in association with two jets in \(pp\) collisions at \(\sqrt{s} = 13\,\text{TeV}\) with the ATLAS detector}",
    journal        = "JHEP",
    volume         = "06",
    year           = "2024",
    pages          = "192",
    doi            = "10.1007/JHEP06(2024)192",
    reportNumber   = "CERN-EP-2024-082",
    eprint         = "2403.15296",
    archivePrefix  = "arXiv",
    primaryClass   = "hep-ex",
}

@Article{STDM-2018-36,
    author         = "{ATLAS Collaboration}",
    title          = "{Measurement of the cross-sections of the electroweak and total production of a \(Z \gamma\) pair in association with two jets in \(pp\) collisions at \(\sqrt{s} = 13\,\text{TeV}\) with the ATLAS detector}",
    journal        = "Phys. Lett. B",
    volume         = "846",
    year           = "2023",
    pages          = "138222",
    doi            = "10.1016/j.physletb.2023.138222",
    reportNumber   = "CERN-EP-2023-098",
    eprint         = "2305.19142",
    archivePrefix  = "arXiv",
    primaryClass   = "hep-ex",
}

@Article{STDM-2018-59,
    author         = "{ATLAS Collaboration}",
    title          = "{Measurement of electroweak \(Z(\nu\bar{\nu})\gamma jj\) production and limits on anomalous quartic gauge couplings in \(pp\) collisions at  \(\sqrt{s} = 13\,\text{TeV}\) with the ATLAS detector}",
    journal        = "JHEP",
    volume         = "06",
    year           = "2023",
    pages          = "082",
    doi            = "10.1007/JHEP06(2023)082",
    reportNumber   = "CERN-EP-2022-138",
    eprint         = "2208.12741",
    archivePrefix  = "arXiv",
    primaryClass   = "hep-ex",
}

@Article{STDM-2019-09,
    author         = "{ATLAS Collaboration}",
    title          = "{Observation of \(WWW\) Production in \(pp\) Collisions at \(\sqrt{s} = 13\,\text{TeV}\) with the ATLAS Detector}",
    journal        = "Phys. Rev. Lett.",
    volume         = "129",
    year           = "2022",
    pages          = "061803",
    doi            = "10.1103/PhysRevLett.129.061803",
    reportNumber   = "CERN-EP-2021-243",
    eprint         = "2201.13045",
    archivePrefix  = "arXiv",
    primaryClass   = "hep-ex",
}

@Article{STDM-2019-17,
    author         = "{ATLAS Collaboration}",
    title          = "{Observation of \(WZ\gamma\) Production in \(pp\) Collisions at \(\sqrt{s} = 13\,\text{TeV}\) with the ATLAS Detector}",
    journal        = "Phys. Rev. Lett.",
    volume         = "132",
    year           = "2024",
    pages          = "021802",
    doi            = "10.1103/PhysRevLett.132.021802",
    reportNumber   = "CERN-EP-2023-095",
    eprint         = "2305.16994",
    archivePrefix  = "arXiv",
    primaryClass   = "hep-ex",
}

@Article{STDM-2021-09,
    author         = "{ATLAS Collaboration}",
    title          = "{Measurement of \(Z\gamma\gamma\) production in \(pp\) collisions at \(\sqrt{s} = 13\,\text{TeV}\) with the ATLAS detector}",
    journal        = "Eur. Phys. J. C",
    volume         = "83",
    year           = "2023",
    pages          = "539",
    doi            = "10.1140/epjc/s10052-023-11579-8",
    reportNumber   = "CERN-EP-2022-192",
    eprint         = "2211.14171",
    archivePrefix  = "arXiv",
    primaryClass   = "hep-ex",
}

@Article{STDM-2022-06,
    author         = "{ATLAS Collaboration}",
    title          = "{Observation of electroweak production of \(W^+W^-\) in association with jets in proton--proton collisions at \(\sqrt{s} = 13\,\text{TeV}\) with the ATLAS Detector}",
    journal        = "JHEP",
    volume         = "07",
    year           = "2024",
    pages          = "254",
    doi            = "10.1007/JHEP07(2024)254",
    reportNumber   = "CERN-EP-2024-015",
    eprint         = "2403.04869",
    archivePrefix  = "arXiv",
    primaryClass   = "hep-ex",
}

@Article{abcd,
	author         = "{ATLAS Collaboration}",
	title = "{Observation of $W^{+}W^{-}\gamma$ production in $pp$ collisions at $\sqrt{s}$ = 13 TeV with the ATLAS detector and constraints on anomalous quartic gauge-boson couplings}",
	year = "2025",
	reportNumber = "CERN-EP-2025-187",
	eprint = "2509.14070",
	archivePrefix  = "arXiv",
	primaryClass   = "hep-ex",
}

@article{Evans:2008zzb,
	editor = "Evans, Lyndon and Bryant, Philip",
	title = "{LHC Machine}",
	doi = "10.1088/1748-0221/3/08/S08001",
	journal = "JINST",
	volume = "3",
	pages = "S08001",
	year = "2008"
}

@article{Lee:1977yc,
	author = "Lee, Benjamin W. and Quigg, C. and Thacker, H. B.",
	title = "{The Strength of Weak Interactions at Very High-Energies and the Higgs Boson Mass}",
	reportNumber = "FERMILAB-PUB-77-022-T",
	doi = "10.1103/PhysRevLett.38.883",
	journal = "Phys. Rev. Lett.",
	volume = "38",
	pages = "883--885",
	year = "1977"
}

@article{Lee:1977eg,
	author = "Lee, Benjamin W. and Quigg, C. and Thacker, H. B.",
	title = "{Weak Interactions at Very High-Energies: The Role of the Higgs Boson Mass}",
	reportNumber = "FERMILAB-PUB-77-030-T",
	doi = "10.1103/PhysRevD.16.1519",
	journal = "Phys. Rev. D",
	volume = "16",
	pages = "1519",
	year = "1977"
}

@article{Eboli:2003nq,
	author = "Eboli, O. J. P. and Gonzalez-Garcia, M. C. and Lietti, S. M.",
	title = "{Bosonic quartic couplings at CERN LHC}",
	eprint = "hep-ph/0310141",
	archivePrefix = "arXiv",
	reportNumber = "YITP-SB-54-03",
	doi = "10.1103/PhysRevD.69.095005",
	journal = "Phys. Rev. D",
	volume = "69",
	pages = "095005",
	year = "2004"
}

@article{Eboli:2006wa,
	author = "Eboli, O. J. P. and Gonzalez-Garcia, M. C. and Mizukoshi, J. K.",
	title = "{$p p \to j j e^{\pm} \mu^{\pm}\nu\nu$ and $j j e^{\pm} \mu^{\mp} \nu \nu$ at $\mathcal{O}(\alpha^{6}_{\text{em}})$ and $\mathcal{O}(\alpha^{4}_{\text{em}}\alpha^{2}_{\text{s}})$ for the study of the quartic electroweak gauge boson vertex at CERN LHC}",
	eprint = "hep-ph/0606118",
	archivePrefix = "arXiv",
	reportNumber = "YITP-SB-06-10, IFUSP-1620-2006",
	doi = "10.1103/PhysRevD.74.073005",
	journal = "Phys. Rev. D",
	volume = "74",
	pages = "073005",
	year = "2006"
}

@article{Chang:2013aya,
	author = "Chang, Jung and Cheung, Kingman and Lu, Chih-Ting and Yuan, Tzu-Chiang",
	title = "{WW scattering in the era of post-Higgs-boson discovery}",
	eprint = "1303.6335",
	archivePrefix = "arXiv",
	primaryClass = "hep-ph",
	doi = "10.1103/PhysRevD.87.093005",
	journal = "Phys. Rev. D",
	volume = "87",
	pages = "093005",
	year = "2013"
}

@article{Espriu:2012ih,
	author = "Espriu, Dom{\`e}nec and Yencho, Brian",
	title = "{Longitudinal WW scattering in light of the {\textquotedblleft}Higgs boson{\textquotedblright} discovery}",
	eprint = "1212.4158",
	archivePrefix = "arXiv",
	primaryClass = "hep-ph",
	reportNumber = "ICCUB-12-480, UB-ECM-PF-84-12",
	doi = "10.1103/PhysRevD.87.055017",
	journal = "Phys. Rev. D",
	volume = "87",
	number = "5",
	pages = "055017",
	year = "2013"
}




\end{document}